\documentclass[12pt]{article}
\usepackage{amsmath,amssymb,mathrsfs,url,graphicx}
\usepackage{color}

\title{Bohmian Mechanics for a Degenerate Time Foliation}

\author{
Ward Struyve\footnote{Department of Physics,
	Universit\'e de Li\`ege,
	B\^atiment B15, Sart Tilman,
	4000 Li\`ege, Belgium. E-mail: ward.struyve@ulg.ac.be}\ \ and 
Roderich Tumulka\footnote{Department of Mathematics,
     Rutgers University, Hill Center, 
     110 Frelinghuysen Road, Piscataway, NJ 08854-8019, 
     USA. E-mail: tumulka@math.rutgers.edu}
}

\date{May 11, 2015}

\newcommand{\be}{\begin{equation}}
\newcommand{\ee}{\end{equation}}
\newcommand{\foliation}{\mathscr{F}} 
\newcommand{\M}{\mathscr{M}} 
\newcommand{\C}{\mathscr{C}} 
\newcommand{\vx}{\boldsymbol{x}}

\renewcommand{\Im}{\mathrm{Im}}
\newcommand{\RRR}{\mathbb{R}}
\newcommand{\CCC}{\mathbb{C}}

\begin{document}
\maketitle
\begin{abstract}
The version of Bohmian mechanics in relativistic space-time that works best, the hypersurface Bohm--Dirac model, assumes a preferred foliation of space-time into spacelike hypersurfaces (called the time foliation) as given. We consider here a degenerate case in which, contrary to the usual definition of a foliation, several leaves of the time foliation have a region in common. That is, if we think of the time foliation as a 1-parameter family of hypersurfaces, with the hypersurfaces moving towards the future as we increase the parameter, a degenerate time foliation is one for which a part of the hypersurface does not move as we increase the parameter. We show that the hypersurface Bohm--Dirac model still works in this situation; that is, we show that a Bohm-type law of motion can still be defined, and that the appropriate $|\psi|^2$ distribution is still equivariant with respect to this law. 

\medskip

\end{abstract}

\section{Introduction}

The natural extension of Bohmian mechanics to relativistic space-time (flat or curved), known as the hypersurface Bohm--Dirac model \cite{HBD}, makes use of a spacelike foliation of space-time into spacelike hypersurfaces, the ``time foliation'' $\foliation$; it can be specified by a 1-parameter family of spacelike hypersurfaces $\Sigma_t$, called ``time leaves,'' with $t$ an arbitrary parameter such that for $t_1\leq t_2$, $\Sigma_{t_2}$ lies in the future of $\Sigma_{t_1}$. In terms of arbitrary space-time coordinates $x^0,x^1,x^2,x^3$, $\Sigma_t$ can be specified by means of a function $f(t,x^1,x^2,x^3)$ as the set
\be\label{Sigmatf}
\Sigma_t=\Bigl\{(x^0,x^1,x^2,x^3)\Big| x^0=f(t,x^1,x^2,x^3)\Bigr\};
\ee
the function $f$ is then increasing in the variable $t$, and the property that $\Sigma_t$ is spacelike corresponds to certain bounds on the derivatives of $f$ with respect to $x^1,x^2,x^3$.

\begin{figure}[h]
\begin{center}
\includegraphics[width=.5 \textwidth]{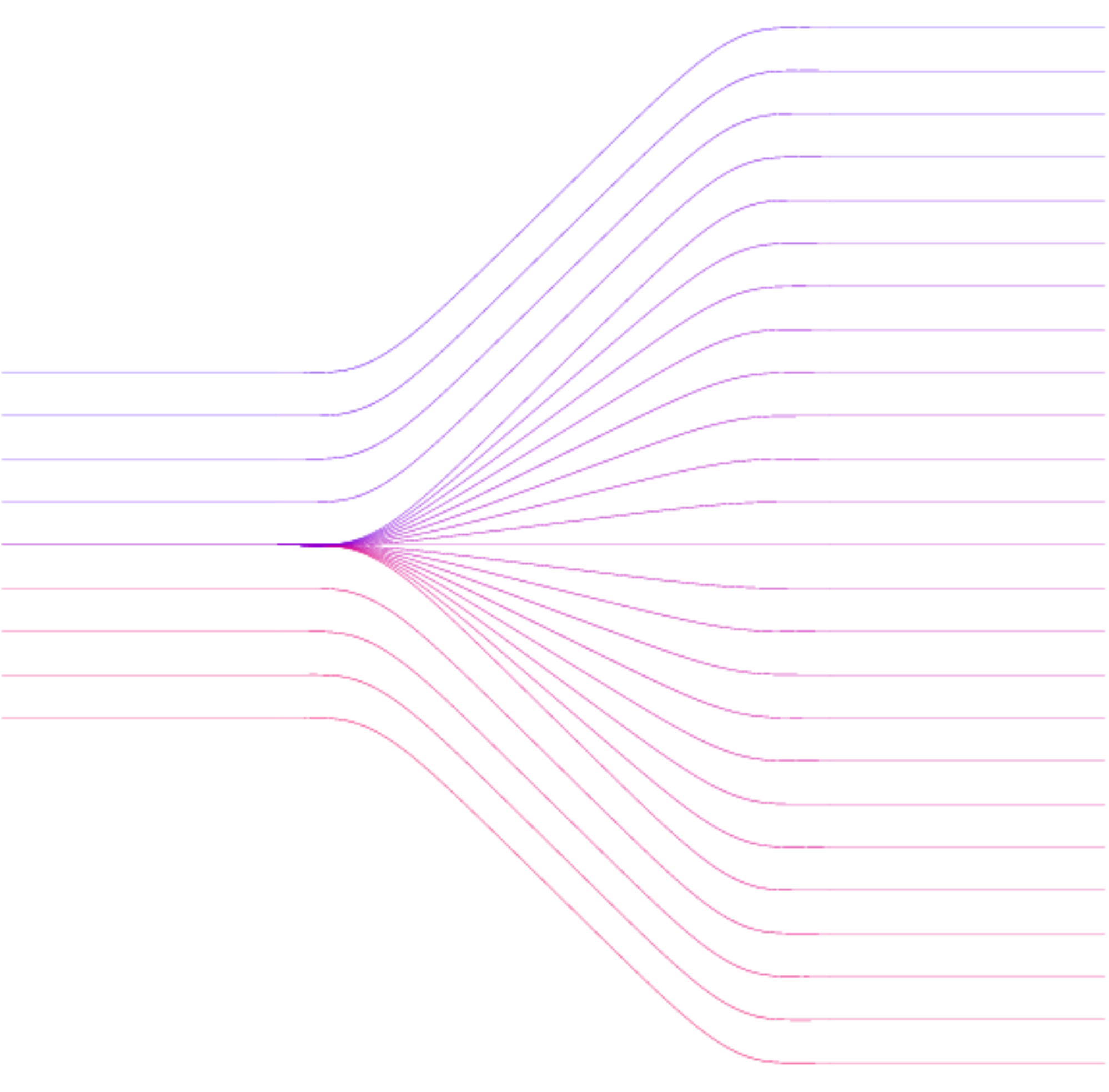}
\end{center}
\caption{An example of a what we mean by a \emph{degenerate} foliation (here, of 1+1-dimensional space-time): some of the leaves overlap in a region (here, on the left). The equations defining this particular example are given in Appendix~\ref{app:ex}.}
\label{figone}
\end{figure}

The degenerate case we are interested in in this note, shown in Figure~\ref{figone}, corresponds to $f$ having a plateau as a function of $t$, i.e., that $f$ is constant as a function of $t$ on some interval $[t_1,t_2]$,
\be
f(t_1,x^1,x^2,x^3) = f(t,x^1,x^2,x^3)=f(t_2,x^1,x^2,x^3)
\ee
for all $t\in [t_1,t_2]$ and all $(x^1,x^2,x^3)$ in some region $A\subset\RRR^3$. In particular,
\be
\frac{\partial f}{\partial t} =0
\ee
for $t\in [t_1,t_2]$ and $(x^1,x^2,x^3)$ in the relevant region $A$. Put differently, when we think of how $\Sigma_t$ moves through space-time as we increase the parameter $t$, we allow that a part of $\Sigma_t$ does not actually move towards the future but remains constant. That is, there is a piece of hypersurface common to all $\Sigma_t$ with $t\in [t_1,t_2]$. Note that such a degenerate foliation cannot arise as the level sets of the 0th coordinate function of a space-time coordinate system because some space-time points $x$ lie on several $\Sigma_t$ (while any coordinate function would yield a unique value at $x$). An explicit example of a degenerate foliation, defined by means of formulas, is provided in Appendix~\ref{app:ex}.

\begin{figure}[h]
\begin{center}
\includegraphics[width=.5 \textwidth]{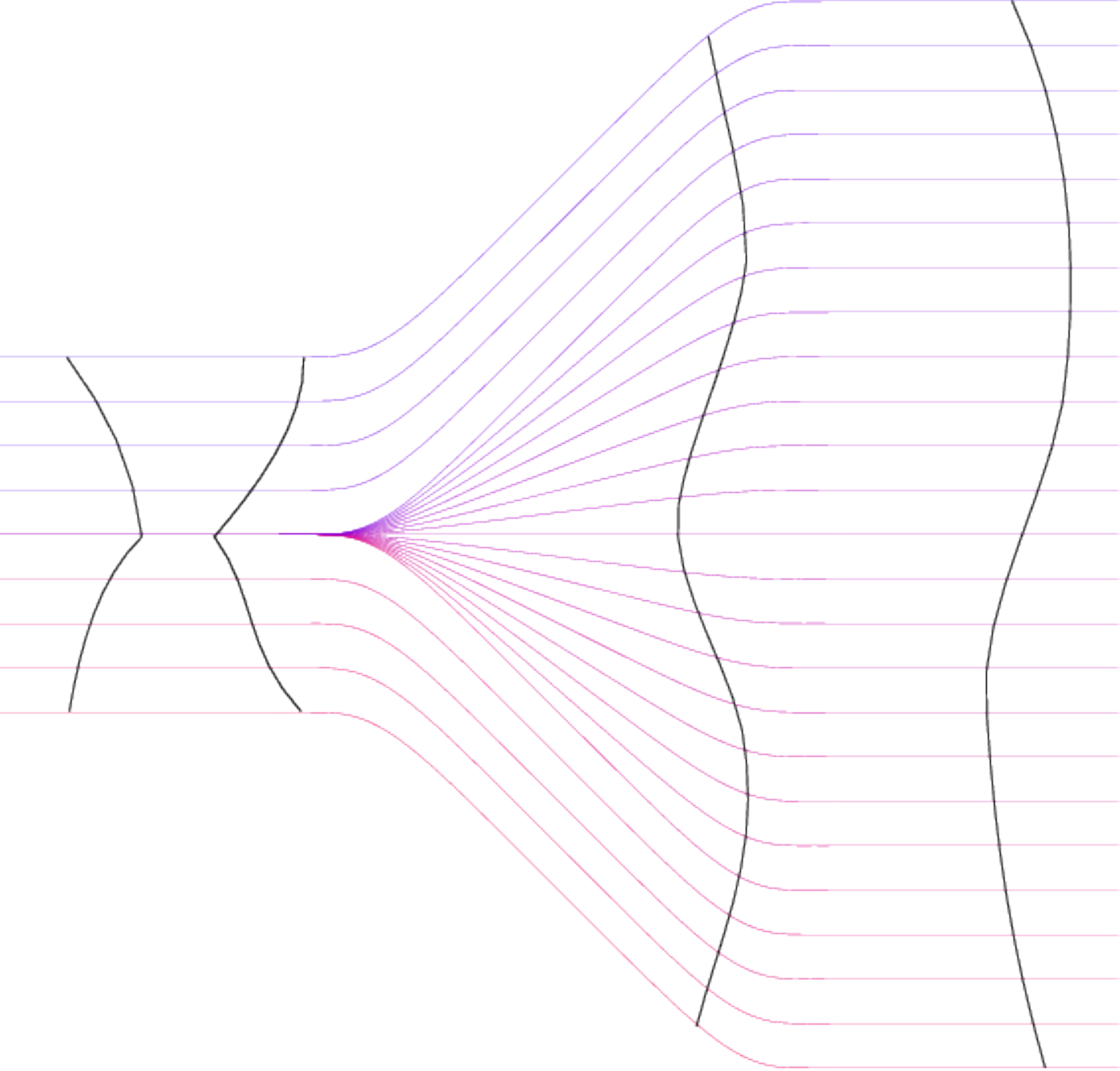}
\end{center}
\caption{The same degenerate foliation as in Figure~\ref{figone}, along with examples of Bohmian world lines. The latter typically have kinks when crossing a leaf in a region where the time foliation is degenerate (i.e., where leaves overlap).}
\label{fig:kink}
\end{figure}

We show here that Bohmian mechanics still works for a time foliation that is degenerate in this sense. More precisely, we show that the definition of the hypersurface Bohm--Dirac model can be extended to this case in such a way that the $|\psi|^2$ distribution is still equivariant. We also note that the Bohmian world lines typically have kinks (jump-like changes of direction) when crossing a plateau of $f$, see Figure~\ref{fig:kink}.

While we are not suggesting that a degenerate time foliation actually occurs in nature, our result is useful for considerations about the flow of probability, which can often be expressed in a particularly intuitive way in terms of Bohmian trajectories. That is, if we think of moving a spacelike hypersurface around in space-time, corresponding to a 1-parameter family $\Sigma_t$, then probability will get transported around on $\Sigma_t^N$ in agreement with the Bohmian motion, \emph{and this is still true if the family $\Sigma_t$ is degenerate}. Thus, our result provides greater freedom in how the hypersurfaces can be moved around, and it is sometimes desirable to push part of the hypersurface to the future while keeping another part unchanged; see \cite{PT15} for an application of this strategy outside of Bohmian mechanics. 

Besides, our result also contributes to illustrating that the hypersurface Bohm--Dirac model is very robust in the sense that it works in many variations of the original setting; previous results in this direction have shown that the hypersurface Bohm--Dirac model also works in curved space-time \cite{Tum01}, in space-times with singularities \cite{Tum08}, and for hypersurfaces $\Sigma_t$ that are not smooth but have kinks \cite{ST14}.

\begin{figure}[h]
\begin{center}
\includegraphics[width=.5 \textwidth,]{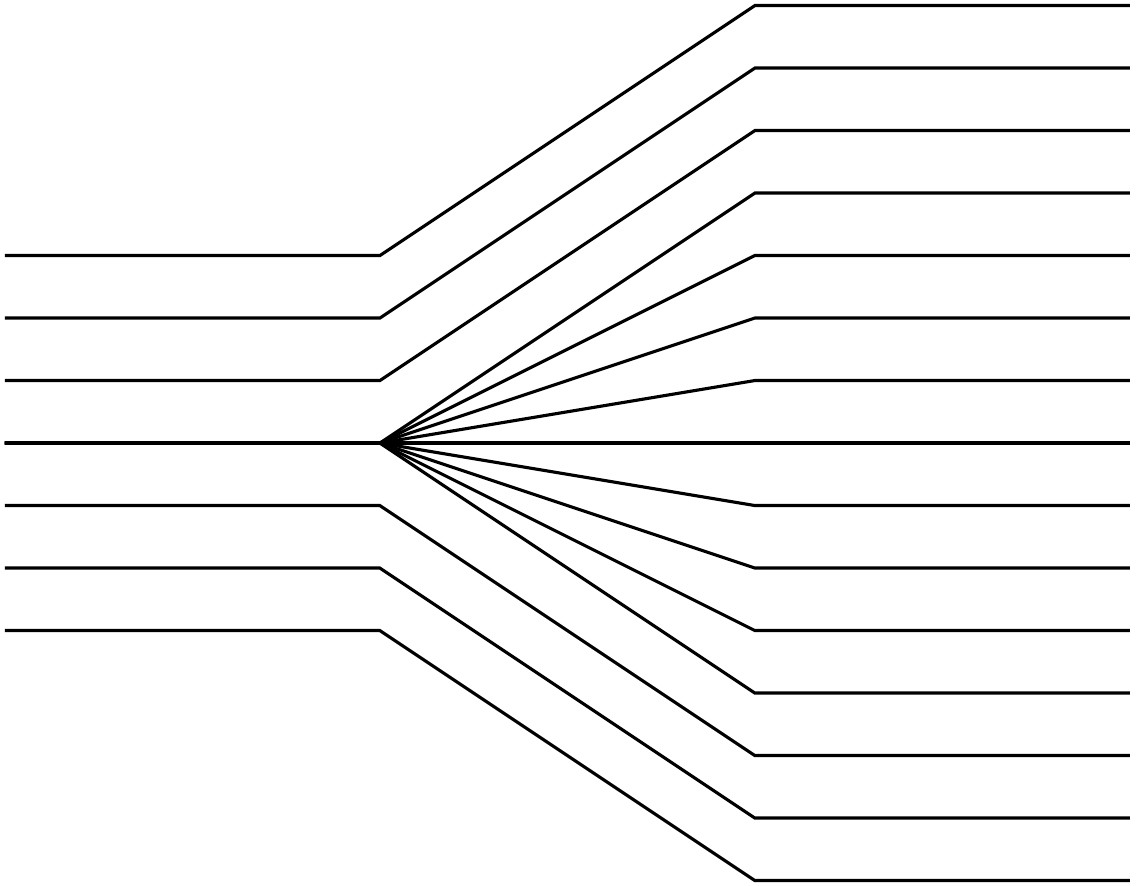}
\end{center}
\caption{Equivariance also holds for this kind of non-smooth, degenerate foliation.}
\label{fig:abrupt}
\end{figure}

Our result also applies to Bohmian theories with a field ontology instead of a particle ontology, and still applies if we drop the assumption $\partial f/\partial t\geq 0$, as we elucidate in Section~\ref{sec:rem}. Moreover, also in the case that $\foliation$ is not everywhere smooth but has kinks, equivariance continues to hold. This phenomenon is discussed in detail in \cite{ST14} for non-degenerate foliations consisting of surfaces $\Sigma_t$ that are manifolds with kinks; the reasons described in \cite{ST14} apply, in fact, equally when $\foliation$ is degenerate and/or when $\foliation$ involves a (continuous but) non-smooth succession of hypersurfaces. As a consequence, equivariance still holds for $\foliation$ of the kind depicted in Figure~\ref{fig:abrupt}.

This note is organized as follows. In Section~\ref{sec:HBD}, we recall the definition of the hypersurface Bohm--Dirac model; in Section~\ref{sec:degenerate}, we extend the definition to the degenerate case and show that equivariance is retained; in Section~\ref{sec:rem}, we collect some remarks; in Appendix~\ref{app:ex}, we provide
an example of a degenerate foliation.

\section{The Hypersurface Bohm--Dirac Model}
\label{sec:HBD}

This model is defined for a non-degenerate (spacelike) time foliation $\foliation=\{\Sigma_t:t\in\RRR\}$ and $N$ particles as follows \cite{HBD}. Let $\M$ denote space-time; readers may take this to be Minkowski space-time, but the model works also for a curved space-time. The wave function $\psi:\M^N \to (\CCC^4)^{\otimes N}$ evolves according to a system of non-interacting multi-time Dirac equations ($c=1=\hbar$)
\be
i\gamma^\mu_k \bigl(\partial_{k\mu}-ieA_\mu(x_k)\bigr) \psi = m_k \psi
\ee
for all $k=1,\ldots,N$, with $\partial_{k\mu} = \partial/\partial x^\mu_k$, $\gamma^\mu_k= 1\otimes \cdots \otimes 1 \otimes \gamma^\mu\otimes 1\otimes \cdots \otimes 1$ with $\gamma^\mu$ in the $k$-th place, and where $A_\mu$ is an external vector potential. Each of the $N$ particles has a world line that is everywhere time- or lightlike (in fact \cite{TT10}, almost everywhere timelike, except in very special situations), whose unique intersection point with $\Sigma_t$ we denote by $X_k(t)$. The law of motion reads
\be\label{motion}
\frac{dX^\mu_k}{dt} \propto j_k^\mu=\Bigl(\overline\psi [\gamma^{\mu_1}\otimes \cdots\otimes \gamma^{\mu_N}] \psi  \Bigr)(X_1(t),\ldots,X_N(t)) \,  \delta^\mu_{\mu_k}\, \prod_{j\neq k} n_{\mu_j}(X_j(t))
\ee
with $n_\mu(x)$ the future-pointing unit normal vector to $\Sigma_t$ at $x\in\Sigma_t$. 

For any spacelike hypersurface $\Sigma$, we say ``the $|\psi|^2$ distribution'' for the probability distribution on $\Sigma^N$ with density, relative to the Riemannian volume measure on $\Sigma^N$ defined by the 3-metric on $\Sigma$, given by
\be
\rho(x_1,\ldots,x_N) = \Bigl(\overline\psi [\gamma^{\mu_1}\otimes \cdots\otimes \gamma^{\mu_N}] \psi  \Bigr)(x_1,\ldots,x_N) \, \prod_{j=1}^N n^\Sigma_{\mu_j}(x_j)
\label{distribution}
\ee
for all $x_1,\ldots,x_N\in \Sigma$, with $n^\Sigma_\mu(x)$ the future-pointing unit normal vector to $\Sigma$ at $x\in\Sigma$. The \emph{equivariance} property of the hypersurface Bohm--Dirac model asserts that if $(X_1(t_0),\ldots,X_N(t_0))$ is $|\psi^2|$-distributed on $\Sigma_{t_0}\in\foliation$, then $(X_1(t),\ldots,X_N(t))$ is $|\psi|^2$-distributed on $\Sigma_t\in\foliation$ for any $t\in\RRR$.

\section{The Hypersurface Bohm--Dirac Model for a Degenerate Time Foliation}
\label{sec:degenerate}

We postulate the following law of motion for the version of the model for a degenerate time foliation $\foliation=\{\Sigma_t:t\in\RRR\}$ specified by a function $f$ as in \eqref{Sigmatf}; that is, we demand that $f(t,\RRR^3)$ is a spacelike hypersurface $\Sigma_t$ and that $\partial f/\partial t \geq 0$. For any $k$ such that $\foliation$ is locally non-degenerate at $X_k(t)$, i.e., such that
\be
\frac{\partial f}{\partial t}(t,X_k^1(t),\ldots,X_k^3(t)) \neq 0\,,
\ee
we keep \eqref{motion} as the law of motion. For any other $k$, i.e., for any $k$ such that $\foliation$ is degenerate at $X_k(t)$, we set
\be\label{motiondeg}
\frac{dX^\mu_k}{dt} =0\,,
\ee
which means that we do not move the point $X_k$ in space-time when $\Sigma_t$ does not move at that point as we increase $t$. This law is the obvious choice, as Eq.~\eqref{motion}, with the proportionality factor made explicit, is of the form
\be\label{motionalt}
\frac{dX^\mu_k}{dt} = (n_\nu j_k^\nu)^{-1} \, n_0\, \partial_t f\, j^\mu_k\,,
\ee 
(with $n_\mu = (1,- {\boldsymbol \nabla} f)/(\sqrt{1- |{\boldsymbol \nabla}f|^2})$), which vanishes when $\partial_t f=0$; thus, \eqref{motiondeg} corresponds to keeping \eqref{motionalt} also at degeneracies. In other words, our law of motion for a degenerate time foliation is a limiting case of the usual law \eqref{motion} for a non-degenerate time foliation. Moreover, \eqref{motiondeg} is the only possible choice (compatible with our parameterization of the world line defined by the relation $X_k(t)\in\Sigma_t$) that leads to world lines that are everywhere time- or lightlike. (That is because, as $t$ increases, we cannot have $X_k(t+dt)$ in the future of $X_k(t)$ if we want it to be on $\Sigma_{t+dt}$, and we cannot have it anywhere on $\Sigma_{t+dt}$ other than at $X_k(t)$ if the curve $X_k(\cdot)$ can never be spacelike.) 

So it is obvious how to choose the law of motion, and the only question is whether this choice leads to equivariance of the $|\psi|^2$ distribution. We will show presently that it does. This result is perhaps not surprising, as the law of motion is a limiting case of the law of motion for the non-degenerate case, which is known to lead to equivariance \cite{HBD,Tum01}. 

\begin{figure}[h]
\begin{center}
\includegraphics[width=7cm]{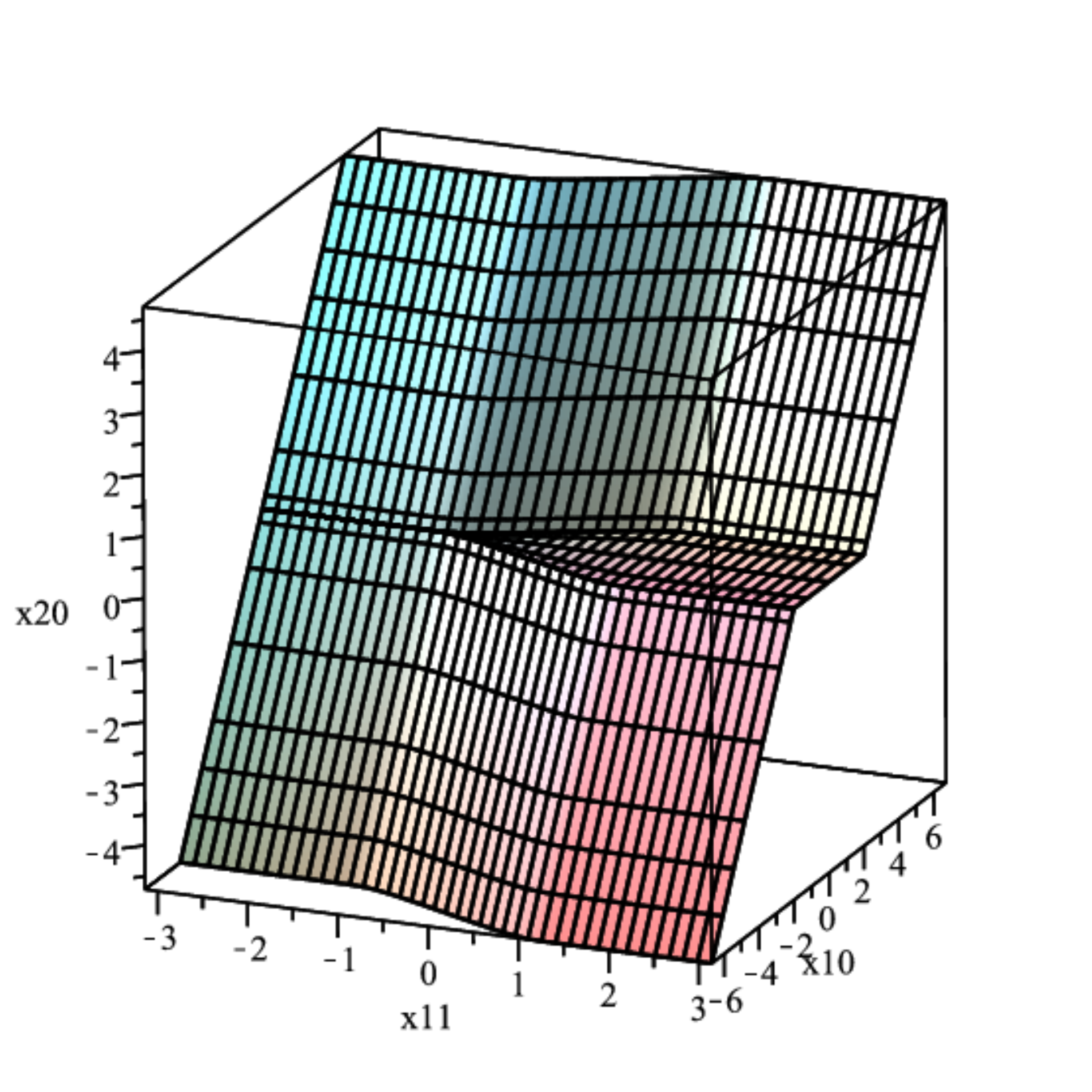}
\ \includegraphics[width=7cm]{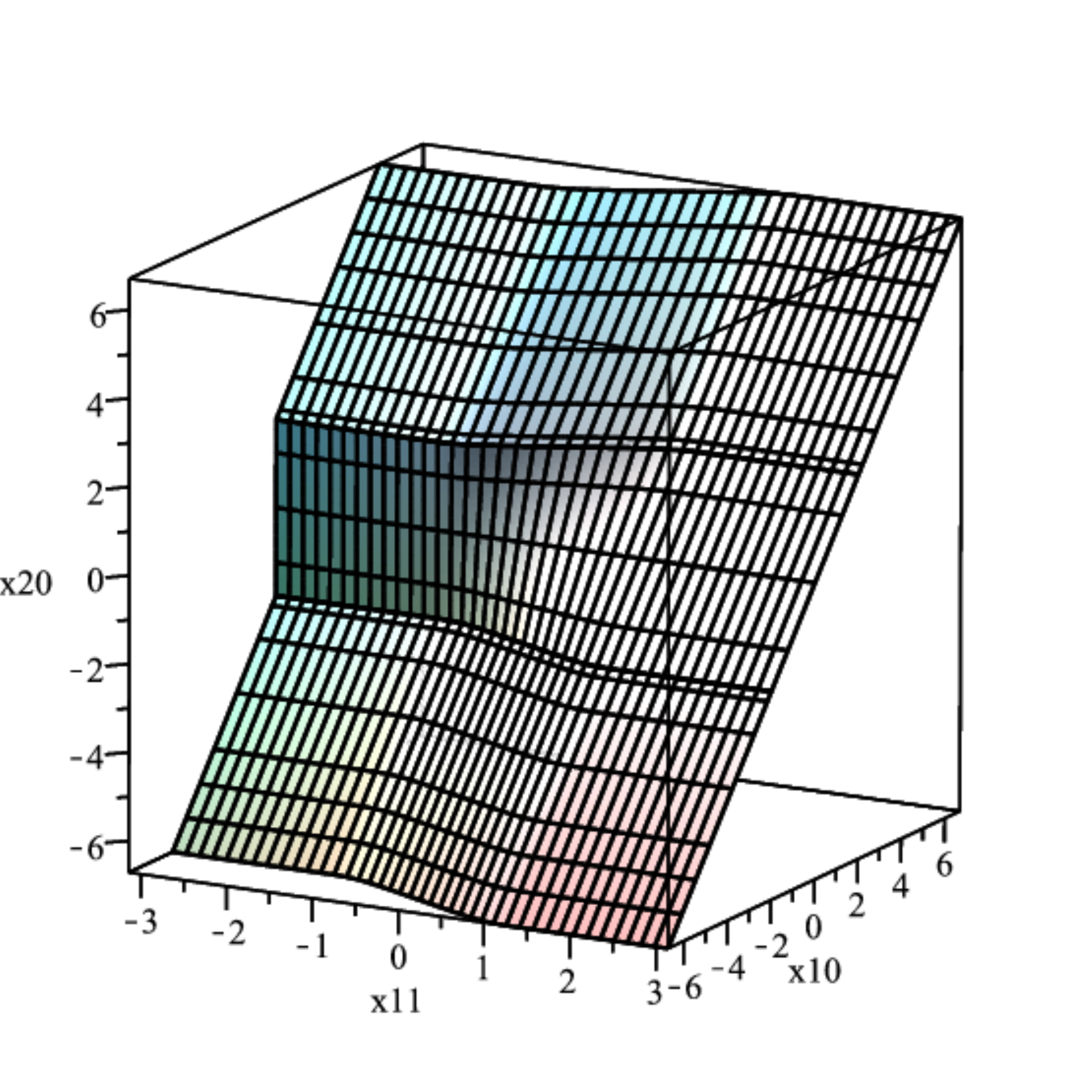}
\end{center}
\caption{Two cross-sections of the surface $\C$ as in \eqref{Cdef} for the particular foliation shown in Figure~\ref{figone} in $1+1$-dimensional space-time for $N=2$ particles; $\C$ is a surface of dimension $(1N+1)=3$ in $\M^N$, which has dimension $(1+1)N=4$. To visualize $\C$, we set the coordinate $x_2^1=\mathrm{const.}$ and display the 2-dimensional surface in 3-dimensional space thus obtained. LEFT: for $x_2^1$ negative and $|x_2^1|$ sufficiently large (in fact, the figure does not depend on the value $x_2^1$ as long as $x_2^1<-\pi/2$). RIGHT: for $x_2^1$ positive and sufficiently large.}
\label{fig:C}
\end{figure}

To verify equivariance, we can revisit the equivariance proofs for the hypersurface Bohm--Dirac model given in \cite{HBD,Tum01} and check that non-degeneracy is not necessary for the proof; we follow here \cite{Tum01}. Suppose that 
\be\label{Cdef}
\C=\bigcup_{t\in\RRR} \Sigma_t^N
\ee
is a piecewise smooth $(3N+1)$-dimensional surface in $\M^N$; see Figure~\ref{fig:C} for an example. (We conjecture that there are degenerate foliations for which $\C$ is smooth rather than merely piecewise smooth, i.e., for which $\C$ has no kinks.{\footnote{We have two reasons for this conjecture. First, it seems that degeneracy, although it violates the standard definition of a foliation, should not disturb the smoothness of $\C$ because degeneracy corresponds to a certain property about the directions tangent to $\C$: If $\foliation$ is degenerate around, say, $\Sigma_t$ at $x_k\in\Sigma_t$, then $T_{(x_1\ldots x_N)} \C \subseteq T_{x_k} \Sigma_t \oplus\bigoplus_{j\neq k} T_{x_j}\M$, where $T_xM$ means the tangent space to the manifold $M$ at the point $x$. The point is that this property has nothing to do with smoothness. Second, we believe that the example in Appendix~\ref{app:ex} can be so modified as to have smooth $\C$ at the expense of greater complexity of the example.}} However, our considerations do not require smoothness and work just as well with kinks.)

The probability current tensor defined by the wave function $\psi$,
\be
j^{\mu_1\ldots \mu_N}(x_1\ldots x_N) = \overline{\psi} [\gamma^{\mu_1}\otimes \cdots \otimes \gamma^{\mu_N}] \psi\,,
\ee
can be transformed into a $3N$-form $J$ \cite{Tum01},
\be\label{Jdef}
J_{\kappa_1\lambda_1\mu_1\ldots \kappa_N\lambda_N\mu_N} = \varepsilon_{\kappa_1\lambda_1\mu_1\nu_1} \cdots \varepsilon_{\kappa_N\lambda_N\mu_N\nu_N} \, j^{\nu_1\ldots \nu_N}\,,
\ee
which is closed on $\M^N$ and thus also on $\C$. Like any $3N$-form on a $(3N+1)$-dimensional manifold, $J$ defines a field of 1-dimensional subspaces $S_{x_1\ldots x_N}$ on $\C$, its kernels, (except at the points where $J=0$, which are the points where $\psi=0$), and the integral curves of $S_{x_1\ldots x_N}$
are exactly the possible trajectories of the configuration on $\C$ as defined in \eqref{motion} and \eqref{motiondeg} above for the degenerate or non-degenerate case. These trajectories satisfy the \emph{wandering condition} \cite{Tum01} (i.e., they are not closed or almost-closed) because they intersect $\Sigma_t^N$ only once for every $t$. As a consequence \cite{Tum01}, $J$ defines a measure $\mu$ on the set of integral curves of the field $S_{x_1\ldots x_N}$ that agrees with the $3N$-form $J$ on any $3N$-surface in $\C$, and thus in particular on $\Sigma_t^N$. If $\C$ has kinks, then integral curves need to be extended across kinks; that is, for an integral curve ending at a point $(x_1,\ldots,x_N)$ on a kink of $\C$, the integral curve on the other side of the kink starting at $(x_1,\ldots,x_N)$ should be regarded as its extension. For almost every kink point, the extension is unique, and $\mu$ is consistently defined on the set of extended integral curves \cite{ST14}. The measure $\mu$ is normalized (i.e., is a probability measure) because $\psi$ is normalized on each of the $\Sigma_t^N$ (because it is normalized on $\Sigma^N$ for any spacelike Cauchy hypersurface $\Sigma$ in $\M$). On surfaces $\Sigma_t^N$, $\mu$ is the ``$|\psi|^2 $ distribution'' \eqref{distribution}, and so this distribution is equivariant. 

To sum up, non-degeneracy is not needed for equivariance.

\section{Remarks}
\label{sec:rem}

\begin{enumerate}
\item Relation to no-signaling. Our result, that equivariance holds also for degenerate foliations, is related to the well-known \emph{no-signaling theorem} as follows. Let $A$ be the region of degeneracy, i.e., the region on the hypersurface $\Sigma_t$ that does not move as we increase $t$ (the left region in Figure~\ref{figone}), and let $A^c_t$ be its complement in $\Sigma_t$ for $t\in[t_1,t_2]$. It is a necessary condition for equivariance that the marginal distribution of $|\psi_{\Sigma_t}|^2$ for the particles in $A$ does not change as we increase $t$. Indeed, if that condition did not hold, the $|\psi|^2$ could not be preserved without moving the particles in $A$. Now that condition is more or less equivalent to the no-signaling theorem: Whatever external fields one experimenter chooses in $A^c_t$ between $\Sigma_{t_1}$ and $\Sigma_{t_2}$, they have no effect on the distribution of, e.g., pointer particles in $A$. Alternatively, the no-signaling property is often expressed by saying that the reduced density matrix $\rho^A_t$ for $A$---obtained from the full quantum state on $\Sigma_t$ by tracing out $A^c_t$---does not depend on $t$ between $t_1$ and $t_2$ for any choice of external fields, $\rho^A_t=\rho^A$. The marginal distribution referred to above is just the distribution with density $\langle q|\rho^A|q \rangle$, where $q$ is any configuration in $A$, i.e., the density is the diagonal of the position representation of $\rho^A$; so the $t$-independence of the marginal distribution follows from the $t$-independence of $\rho^A$.

\item Kinks in the world lines (see Figure~\ref{fig:kink}). While the \emph{positions} of the particles in the region $A$ of degeneracy do not change as we increase $t$ from $t_1$ to $t_2$, the \emph{velocities} may well (and will typically) change, in the sense that $j_k^\mu$ as in \eqref{motion} changes with $t$ for particle $k$ in $A$ (unless no particle outside $A$ is entangled with particle $k$), and, as a consequence, 
\be
\lim_{t\searrow t_2} \biggl[ \frac{dX_k^\mu}{dt} \biggr] 
\neq \lim_{t\nearrow t_1} \biggl[\frac{dX_k^\mu}{dt} \biggr]\,,
\ee
where $[v^\mu]$ denotes the 1-dimensional subspace through $v^\mu$. (Equivalently, this relation is also true if $[v^\mu]$ denotes the unit vector in the direction of $v^\mu$, $[v^\mu] =(v_\nu v^\nu)^{-1/2} v^\mu$, provided that the limiting direction is not lightlike.) This means that the world line of particle $k$ has a kink at the point $X$ where it crosses $A$, with both directions (into $X$ and out of $X$) being timelike or lightlike. Particles outside $A$ (i.e., that do not pass through the degeneracy) do \emph{not} feature kinks.

\item No entering or leaving $A$. Since particles in $A$ do not move, they obviously cannot leave $A$ before $t_2$. Conversely, no particle that is outside $A$ on $\Sigma_{t_1}$ can enter $A$ between $t_1$ and $t_2$, as follows from the fact that the world lines are everywhere time- or lightlike.

\item Particle creation and annihilation. We expect that the result of this paper, equivariance of $|\psi|^2$ also for degenerate foliations, extends to ``Bell-type quantum field theories'' \cite{Bell86,crlet,crea2B}, i.e., versions of Bohmian mechanics involving particle creation and annihilation by means of stochastic jumps of the actual configuration in the configuration space of a variable number of particles. These versions are formulated in \cite{Bell86,crlet,crea2B} for a fixed Lorentz frame in flat space-time. It should be straightforward to adapt them to a non-flat time foliation (also in curved space-time), provided the Hamiltonian can be defined for such a foliation; this will be possible 
in a natural way for an arbitrary foliation if a multi-time evolution law for the wave function on the set of spacelike configurations is provided. Such laws do not get along with an ultraviolet cut-off; on the formal level, such a law is described and discussed in \cite{PT13}, while on the rigorous level, it may be possible to formulate such a law using interior--boundary conditions \cite{TT15}. The reason why these theories should work also with degenerate foliations is the following:
They have a law specifying the jump rate in terms of the wave function and the interaction Hamiltonian, where ``rate'' means probability per time and should therefore be proportional to the thickness of the layer between $\Sigma_t$ and $\Sigma_{t+dt}$ at the point where the particle creation or annihilation occurs. For a degenerate foliation, this thickness will sometimes vanish, and therefore no creation or annihilation events should occur at such space-time locations.

\item Field ontologies. There also exist Bohmian approaches with fields as the actual configurations, rather than particle positions (see, e.g., \cite{S10}). These approaches could similarly be formulated with respect to a degenerate foliation. For example, for a scalar field, the guidance equation for a non-degenerate time foliation is of the form \cite{DGNSZ14}
\begin{equation}
\frac{d \varphi (x)}{d\tau} =  \frac{1}{\sqrt{h}}\text{Im} \left( \frac{1}{\Psi_{\Sigma_x}} \frac{\delta \Psi_{\Sigma_x}}{\delta \varphi_{\Sigma_x}(x)} \right)\Big|_{\varphi|_{\Sigma_x}} \,,
\label{fieldguidance}
\end{equation}
where $d \varphi (x) /d\tau = n^\mu(x) \partial_\mu \varphi (x)$ is the directional derivative at $x$ along the normal to the time leaf $\Sigma_x$ that contains $x$, $h$ is the determinant of the induced Riemannian metric on $\Sigma_x$, $\varphi|_{\Sigma_x}$ is the restriction of the field configuration to the hypersurface $\Sigma_x$, and $\Psi_{\Sigma}(\varphi_\Sigma) =\langle \varphi_\Sigma | \Psi \rangle$, where $|\Psi \rangle$ is the state vector in the Heisenberg picture and where $|\varphi_\Sigma\rangle$ is defined by $ \varphi(x) |\varphi_\Sigma\rangle = \varphi_\Sigma(x) |\varphi_\Sigma\rangle$, for points $x$ on $\Sigma$, with $\varphi(x)$ the Heisenberg field operator. In terms of the parameter $t$ of the time leaves $\Sigma_t$, \eqref{fieldguidance} can be rewritten as
\be\label{fieldt}
\frac{\partial \varphi (f(t,\vx),\vx)}{\partial t} =  \frac{\partial f}{\partial t} \biggl[ n_0\frac{1}{\sqrt{h}} \Im \left( \frac{1}{\Psi_{\Sigma_t}} \frac{\delta \Psi_{\Sigma_t}}{\delta \varphi_{\Sigma_t}(f(t,\vx),\vx)} \right)\Big|_{\varphi|_{\Sigma_t}} + u^\mu\partial_\mu \varphi \biggr]\,,
\ee
with $\vx=(x^1,x^2,x^3)$ and $u^\mu$ the orthogonal projection of the timelike coordinate vector $(1,0,0,0)$ to the tangent plane to $\Sigma_t$ at $x=(f(t,\vx),\vx)$, which can be written as $u^\mu=\delta_0^\mu-n_0n^\mu$; note that $u^\mu\partial_\mu\varphi$ is a directional derivative along $\Sigma_t$ and thus known if $\varphi_{\Sigma_t}$ is given.\footnote{To verify \eqref{fieldt}, note that $\partial_t \varphi(f(t,\vx),\vx))= \partial_\mu \varphi  \,\delta_0^\mu \,\partial_t f$, and $\delta_0^\mu=(1,0,0,0)=n_0 n^\mu + u^\mu$.}
If we take \eqref{fieldt} as the guidance equation for $\varphi$ also in the case of a degenerate foliation, it implies that, at any point $x=(f(t,\vx),\vx)$ where the foliation is degenerate (i.e., $\partial f/\partial t=0$),
\be
\frac{\partial \varphi(f(t,\vx),\vx)}{\partial t}=0\,,
\ee
in analogy to \eqref{motiondeg}. We expect that equivariance still holds, for the same reasons as for the particle ontology.

\begin{figure}[h]
\begin{center}
\includegraphics{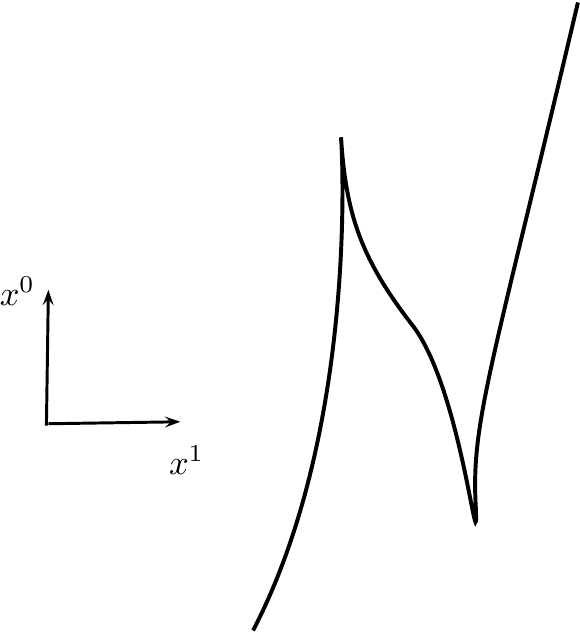}
\end{center}
\caption{The kind of world lines arising if we allow $\Sigma_t$ also to move backwards in time, as explained in Remark~\ref{rem:back}.}
\label{fig:revert}
\end{figure}

\item\label{rem:back} Generalization to a time foliation going backwards in time. We may even drop the assumption $\partial f/\partial t\geq 0$ and allow any smooth $f$ function such that $f(t,\RRR^3)$ is a spacelike hypersurface for every $t$. That means that, as we increase $t$, $\Sigma_t$ may move towards the future in some regions, towards the past in others, and remain unchanged elsewhere, with the regions changing with $t$. Again, we are not suggesting that a time foliation like that actually occurs in nature, but it may be worthwhile noticing this mathematical possibility. The law of motion \eqref{motion} can naturally be adapted to this case by writing it in the form \eqref{motionalt}, which allows for $\partial_t f$ to be positive, negative, or zero. When a world line reaches a point at which $\partial_t f$ changes sign, the world line will typically change sign of its direction, as depicted in Figure~\ref{fig:revert}---an extreme kind of kink.

In this scenario, there is no longer a unique point of intersection between a world line and a time leaf $\Sigma_t$; however, the $4N$ functions $t\mapsto X_k^\mu(t)$ that together form a (smooth!)\ solution of the ODE system \eqref{motionalt} provide an unambiguous choice of $N$ points on $\Sigma_t$ for each $t$. For this configuration, the $|\psi|^2$ distribution is again equivariant. Indeed, consider, instead of $\C$, the $(3N+1)$-dimensional surface
\be
\tilde\C = \bigcup_{t\in\RRR}\{t\}\times \Sigma_t^N
\ee
in the $(4N+1)$-dimensional space $\RRR\times \M^N$ (which is smooth if $f$ is), and on it the closed $3N$-form $J$ defined as in \eqref{Jdef}. Then the kernels of $J$ form again a field of 1-dimensional subspaces on $\tilde\C$, the integral curves of which are the solutions of the law of motion \eqref{motionalt}. Since the $t$ variable is increasing along the integral curves, they obey the wandering condition, and thus \cite{Tum01} the measure defined by $J$ on $\Sigma_t^N$ (i.e., the $|\psi|^2$ distribution) is equivariant.

\end{enumerate}

\appendix

\section{Example of a Degenerate Foliation}
\label{app:ex}

Here is a specific example of an $f$ function such that the foliation it defines according to \eqref{Sigmatf} is degenerate. In fact, this example is depicted in Figures \ref{figone}, \ref{fig:kink}, and \ref{fig:C}.

\begin{figure}[h]
\begin{center}
\includegraphics[width=.7 \textwidth]{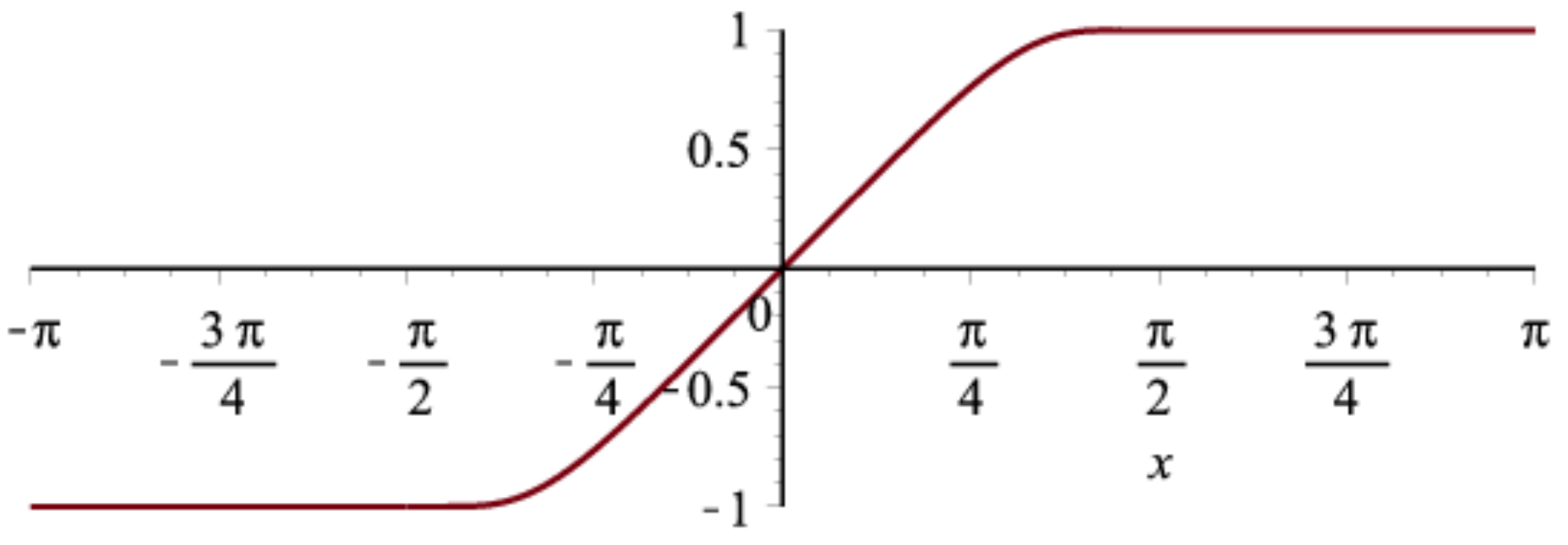}
\end{center}
\caption{Graph of the function $g$ defined in \eqref{gdef}.}
\label{fig:tanhtan}
\end{figure}

Let $a>0$, and let $g:\RRR\to\RRR$ be a smooth increasing function such that $g(x)=-1$ for $x<-a$, $g(x)=1$ for $x>a$, and $dg/dx\leq 1$ everywhere; for example (depicted in Figure~\ref{fig:tanhtan}), $a=\pi/2$ and
\be\label{gdef}
g(x) = \begin{cases} 
-1 & \text{if }x\leq -\pi/2\\
\tanh(\tan(x)) & \text{if } -\pi/2<x<\pi/2\\
1 & \text{if }\pi/2\leq x\,. \end{cases}
\ee

\begin{figure}[h]
\begin{center}
\includegraphics[width=.7 \textwidth]{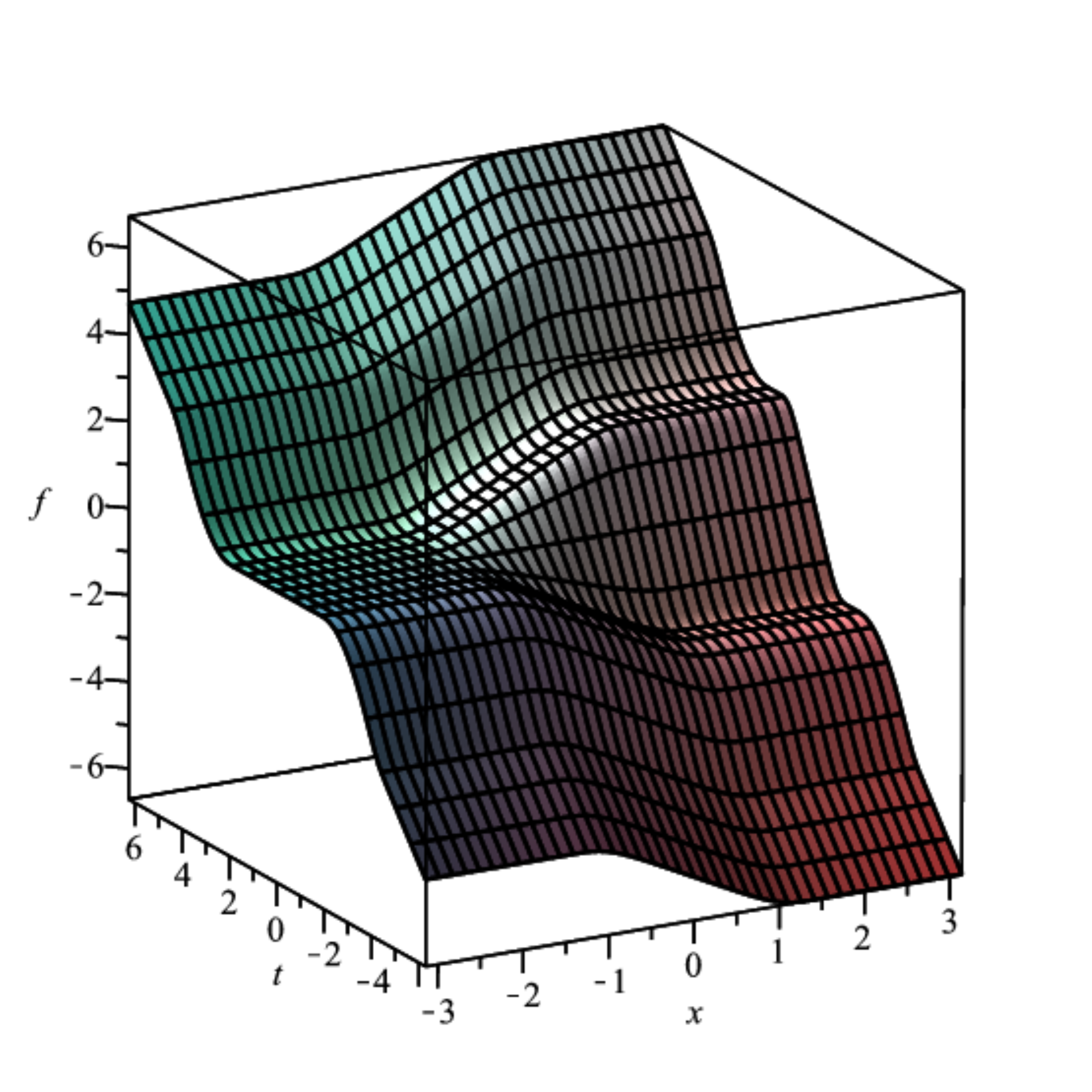}
\end{center}
\caption{Graph of the function $f$ defined in \eqref{fdef}. The plateau $f(t,x)=0$ is visible on the left (for $-\pi/2<t<\pi/2$ and $x<-\pi/2$).}
\label{fig:f}
\end{figure}

Then we define $f(t,x^1,x^2,x^3)$ as follows; the function is actually independent of $x^2$ and $x^3$; for simplicity, we write $x$ instead of $x^1$; $f(t,x)$ is given by
\be\label{fdef}
f(t,x) =\begin{cases}
(t+a)\tfrac{1}{2}\bigl(1-g(t+2a)\bigr)-1-g(x)&\text{if }t\leq -a\\
g(t)(1+g(x))&\text{if }-a\leq t\leq a\\
(t-a)\tfrac{1}{2}\bigl(1+g(t-2a)\bigr) +1+g(x)&\text{if }a\leq t\,.
\end{cases}
\ee
The graph of $f$ is shown in Figure~\ref{fig:f}. Note the plateau, $f(t,x)=0$ for all $(t,x)$ with $-a\leq t \leq a$ and $x\leq -a$. 

Alternatively, if we define
\be\label{fdef2}
f_2(t,x) =\begin{cases}
t+1-g(x)&\text{if }t\leq -2\\
\tfrac{1}{2}t(1+g(x))&\text{if }-2\leq t\leq 2\\
t-1+g(x)&\text{if }2\leq t\,,
\end{cases}
\ee
then we actually obtain the same foliation parameterized differently, i.e., the hypersurfaces are labeled differently; the labels are chosen such that $f_2(t,x)=t$ for $x>a$. The function $f_2$ is not smooth, but the definition is simpler.

The corresponding surface $\C$ is not smooth. For two particles, this can be seen as follows. First note that if $x^1_2 > \pi/2$, then $f_2(t,x^1_2) = t$. As such the intersection of $\C$ with a constant $x^1_2$ hyperplane, which we denote by $\C_{x^1_2}$, is given by
\be
\C_{x^1_2} = \bigcup_{t,x^1_1\in\RRR} (f_2(t,x^1_1),x^1_1,t,x^1_2) = \bigcup_{x^1_1,x^0_2\in\RRR} (f_2(x^0_2,x^1_1),x^1_1,x^0_2,x^1_2) \,.
\ee
Since $\partial f_2(t,x)/\partial t$ is not continuous for $t = \pm 2$ and $x < \pi/2$, $\C_{x^1_2}$ and hence $\C$ are not smooth. This is illustrated in Figure~\ref{fig:C} on the right.

\bigskip

\noindent\textit{Acknowledgments.} 
Both authors acknowledge support from the John Templeton Foundation, grant no.\ 37433. W.S.\ acknowledges current support from the Actions de Recherches Concert\'ees (ARC) of the Belgium Wallonia-Brussels Federation under contract No.\ 12-17/02.

\end{document}